\documentclass[aps,prc,twocolumn,groupedaddress,showpacs,superscriptaddress,nofootinbib]{revtex4-2}
\usepackage{graphicx,amsmath}
\usepackage{amssymb}
\usepackage{amsmath}
\usepackage{booktabs}
\usepackage{tablefootnote}
\usepackage{units}

\usepackage[dvipsnames]{xcolor}

\usepackage{stix}
\usepackage{tikz}
\usepackage{threeparttable,lipsum}
\usepackage{mathtools}
\usepackage{hyperref}
\usepackage{dcolumn}
\usepackage{bm}

\begin{document}


\title{Laboratory electron screening in nuclear resonant reactions}


\author{C. Iliadis}
\email[]{iliadis@unc.edu}
\affiliation{Department of Physics \& Astronomy, University of North Carolina at Chapel Hill, NC 27599-3255, USA}
\affiliation{Triangle Universities Nuclear Laboratory (TUNL), Duke University, Durham, North Carolina 27708, USA}


\date{\today}

\begin{abstract}
Both nonresonant and resonance reaction data are subject to laboratory electron screening effects. For nonresonant reactions, such effects are well documented and the measured cross sections can be corrected to find the unscreened ones. Frequently, the procedure and expression to calculate laboratory electron screening factors for nonresonant reactions are also applied to isolated narrow resonances, without much theoretical support or experimental evidence. 

A simple model is applied to estimate electron screening factors, lengths, and potentials for narrow resonances. The corrections to the measured data result in an {\it enhancement} of the unscreened resonance strengths by less than 0.2\%, contrary to published narrow-resonance screening correction factors, which predict a {\it reduction} of the unscreened strengths by up to 25\%. Unless it can be proven otherwise, it is recommended that measured strengths of isolated narrow resonances not be corrected for laboratory electron screening. 

The prospects of investigating laboratory electron screening effects by measuring almost negligible differences in resonance {\it strengths} are not promising. Instead, the difference of the resonance {\it energy} for the unscreened and screened situation may be measurable. As an example, the case of the $E_r$ $=$ $956$-keV resonance in the $^{27}$Al(p,$\gamma$)$^{28}$Si reaction is discussed. It is also demonstrated that the claim of a previously reported detection of a resonance near $800$~keV in the $^{176}$Lu(p,n)$^{176}$Hf reaction is incorrect. 
\end{abstract}


\maketitle


\section{Introduction}\label{sec:intro}
Nonresonant charged-particle nuclear reaction measurements at low bombarding energies are impacted by the presence of electrons in the vicinity of the interacting nuclei. These electrons, either bound to individual target or projectile atoms, or freely moving in the conduction band in the case of a metal, give rise to an attractive potential that effectively reduces the width of the overall potential barrier to be penetrated by the projectile. Therefore, the astrophysical $S$ factor extracted from a nonresonant cross section measured in the laboratory is expected to be larger compared to the $S$ factor that would have been obtained in the absence of electrons, especially at the lowest bombarding energies. This effect has been observed in several experiments (see, e.g., Ref.~\cite{ALIOTTA2001790}). It is important to correct the measured cross section for such {\it laboratory} electron screening effects, and, thereby, determine the cross section applicable to bare interacting nuclei. The latter quantity can then be used, together with a prescription of {\it stellar} electron screening, to calculate thermonuclear reaction rates, which are an essential ingredient for models of stellar evolution and explosion. The electron screening correction factors differ for the laboratory and stellar environment.  The focus of the present work is on the former. The latter have been calculated, e.g., by Refs.~\cite{Salpeter1954,Mitler}.

Many authors (see e.g., Refs.~ \cite{spitaleri,aliottalanganke}, and references therein) have pointed out that the magnitude of the laboratory electron screening corrections extracted from low-energy nonresonant cross section data are larger than what is predicted from theory. Sophisticated theoretical models have been applied to the problem, but significant inconsistencies between theory and experiment remain (for a review, see Ref.~\cite{aliottalanganke}). 

The aim of the present work is not to provide more accurate predictions for the nonresonant laboratory electron screening corrections, but to investigate the  correction pertaining to isolated narrow resonances. Assenbaum et al. \cite{Assenbaum87} were first to suggest that the electron screening correction factors obtained for nonresonant reactions can be applied equally to narrow resonances. They also predicted that electron screening effects would result in a shift of the resonance energy compared to the case of unscreened nuclei. As will be discussed below, their first claim turns out to be incorrect, while the second one is confirmed in the present work. Measuring such shifts of the resonance energy may allow for a detailed study of the interplay between atomic and nuclear processes. 

The effects of atomic electrons on nuclear resonance {\it scattering} have been studied many times before \cite{BENN1967296,PhysRevLett.45.703,BRIGGS1981436,Heinz_1987}. However, a review of such effects in nuclear resonance {\it reactions} has not been given in any detail. For this reason, in the literature, the correction factors obtained for nonresonant reactions are also applied to narrow resonances (see, e.g., Refs.~\cite{PhysRevC.94.055804,PhysRevLett.117.142502,2012PhLB..707...60S,Sergi2015}). Such corrections always result in a bare (unscreened) resonance strength that is lower, by up to 25\%, depending on the reaction, compared to the measured (screened) strength. However, it is neither obvious why the same laboratory screening correction factors should be applied to both nonresonant and narrow-resonance reaction data, nor whether there are compelling reasons to correct the latter data for laboratory screening effects at all.

In Secs.~\ref{sec:nonres} and \ref{sec:narrow}, laboratory electron screening effects for nonresonant reactions and narrow resonances, respectively, will be reviewed. Screening energies and lengths are presented in Sec.~\ref{sec:screen2}. Results are provided in Sec.~\ref{sec:results} and future measurements are discussed in Sec.~\ref{sec:shift}. A concluding summary is given in Sec.~\ref{sec:summary}.

\section{Electron screening in nonresonant reactions}\label{sec:nonres}
The nonresonant cross section, $\sigma(E)$, at a center-of-mass energy, $E$, can be parameterized as \cite{Iliadis_2015}
\begin{equation}
\label{eq:sfac}
\sigma(E) \equiv \frac{1}{E} S(E) e^{-2 \pi \eta(E)}
\end{equation}
where the astrophysical $S$ factor is frequently a function that varies slowly with energy; $\eta(E)$ denotes the Sommerfeld parameter, $\eta$ $\equiv$ ($Z_0 Z_1 e^2 / \hbar) \sqrt{\mu /(2E)}$; $Z_0$, $Z_1$, $e$, and $\mu$ are the charges of the interacting nuclei, elementary charge, and reduced mass, respectively. The energy-dependent Gamow factor, $e^{- 2 \pi \eta}$, describes the $s$-wave transmission through the Coulomb barrier.

\begin{figure}[ht]
\includegraphics[width=1.0\linewidth]{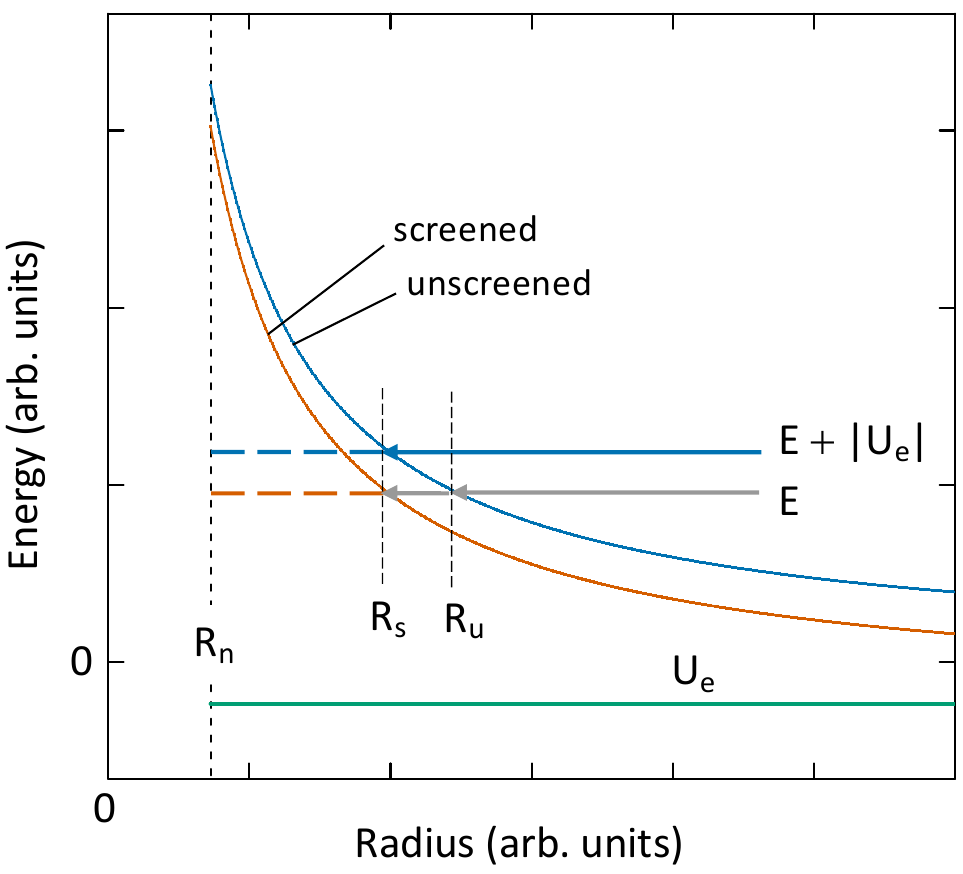}
\caption{\label{fig:nonr} 
Schematic representation (not to scale) of electron screening for a nonresonant charged-particle nuclear reaction in the laboratory, showing the unscreened Coulomb potential (blue curve), constant negative screening potential, $U_e$ (green line), screened Coulomb potential (red curve), total energy, $E$ (gray arrows), and effective energy, $E_{\mathrm{eff}}$ $=$ $E$ $+$ $|U_e|$ (blue arrow); $R_n$, $R_s$, and $R_u$ denote the nuclear radius, and the classical turning points at energy $E$ for the screened and unscreened barrier, respectively. The actual reaction in the laboratory is represented by the second gray arrow (on the left) extending to the red curve. Notice that $R_s$ is also equal to the classical turning point for the unscreened barrier (blue curve) at the effective energy, $E_{\mathrm{eff}}$ (blue arrow). No screening potential is shown inside the nucleus, because it is irrelevant for the derivation of $f_{nr}$ in Eq.~(\ref{eq:nrscreen5}).
}
\end{figure}

The situation is depicted in Fig.~\ref{fig:nonr}. The unscreened Coulomb barrier, $V_C(r)$, is shown as the blue curve. A negative screening potential, $U_e$, is represented by the green line. It is depicted here as a constant potential, which is the usual assumption made in the literature for nonresonant reactions. The magnitude of $U_e$ is highly exaggerated for illustrative purposes. The screened Coulomb potential, $V_C(r)$ $+$ $U_e$, i.e., the sum of the blue and green lines, is shown as the red curve. When a particle is incident on the unscreened barrier at a center-of-mass energy, $E$ (gray arrow at right), it needs to tunnel through a distance $R_u$ $-$ $R_n$ to initiate the reaction, where $R_u$ and $R_n$ denote the classical turning point for the unscreened barrier and the nuclear radius, respectively. A particle of energy $E$ incident on the screened barrier will tunnel through a shorter distance of $R_s$ $-$ $R_n$, where $R_s$ is the classical turning point of the screened barrier. The increase in the measured nonresonant cross section is described by the ratio of transmission probabilities, $T^\prime$ and $T$, through the screened and unscreened barriers, respectively, at energy, $E$,
\begin{equation}
\label{eq:nrscreen2}
f_{nr} \equiv \frac{\sigma_{\mathrm{screen}}}{\sigma_{\mathrm{unscreen}}} = \frac{T^{\prime}(E)}{T(E)}
\end{equation}
The transmission coefficient in the Wentzel-Kramers-Brillouin (WKB) approximation for the unscreened Coulomb barrier is given by \cite{Merzbacher}
\begin{equation}
\label{eq:nrscreen3}
T(E) \approx \exp \left( - \frac{\sqrt{8 \mu}}{\hbar} \int_{R_n}^{R_u} \sqrt{ V_C(r) - E} \, dr \right)
\end{equation}
where $\mu$ is the reduced mass, and $V_C(r)$ is the (unscreened) Coulomb potential. The outer turning point is given by $R_u$ $=$ $Z_0 Z_1 e^2 / (4 \pi \epsilon_0 E) $, with $\epsilon_0$ denoting the vacuum permittivity. For a particle approaching the screened barrier at energy $E$, we can write
\begin{equation}
\label{eq:nrscreen4}
T^\prime(E) \approx \exp \left( - \frac{\sqrt{8 \mu}}{\hbar} \int_{R_n}^{R_u} \sqrt{ V_C(r) + U_e - E } \, dr \right)
\end{equation}
It can be seen that Eq.~(\ref{eq:nrscreen4}) is equivalent to the transmission of the {\it unscreened} barrier at an energy of $E_{\mathrm{eff}}$ $=$ $E$ $+$ $| U_e |$, i.e., $T^\prime(E)$ $=$ $T(E_{\mathrm{eff}})$, as indicated by the blue arrow in Fig.~\ref{fig:nonr}. This is the reason why usually the transmission coefficients, $T^\prime(E)$ and $T(E)$, are not computed numerically. Instead, they are approximated by the Gamow factors, $T(E)$ $\approx$ $\mathrm{exp}(- 2 \pi \eta (E))$ and $T^\prime(E)$ $\approx$ $\mathrm{exp}(- 2 \pi \eta (E_{\mathrm{eff}}))$, so that the nonresonant electron screening correction factor becomes
\begin{equation}
\label{eq:nrscreen5}
f_{nr} \approx \frac{e^{-2 \pi \eta(E_{\mathrm{eff}})}}{e^{-2 \pi \eta(E)}} \approx e^{\pi \eta(E) \frac{|U_e|}{E}}
\end{equation}
In the last step, it is assumed that the energy of the incident particle is large compared to the screening energy, i.e., $E$ $\gg$ $|U_e|$. The electron screening potential, $U_e$, is assumed to be independent of energy. The factor, $f_{nr}$, amounts to unity at higher energies, where $E$ $\gg$ $|U_e|$, and increases as the energy decreases. Therefore, its magnitude is $f_{nr}$ $\ge$ $1$.

Equation~(\ref{eq:nrscreen5}) has been applied in Refs.~\cite{erma,Assenbaum87} and is the commonly adopted formalism for nonresonant cross sections. As can be seen from the above derivation, the incident particle does not actually gain total energy, as is sometimes stated. Instead, the energy shift, from $E$ to $E_{\mathrm{eff}}$, facilitates the convenient calculation of $f_{nr}$ by using the Gamow factors at these two energies (see also Ref.~\cite{Huke2008}), without the need of computing the ratio of transmission coefficients at energy $E$ numerically. Also, sometimes a pre-factor containing the ratio of energies and $S$ factors at $E$ and $E_{\mathrm{eff}}$ is included in Eq.~(\ref{eq:nrscreen5}). This is incorrect since the reaction takes place at energy $E$, not at $E_{\mathrm{eff}}$. 

The electron screening potential for nonresonant reactions can be estimated with a suitable model representing the electronic configuration of the target and projectile. For example, for gaseous targets and low bombarding energies, the adiabatic (limit) approximation is frequently used \cite{Assenbaum87}. It assumes that the electron velocities are much larger than the relative motion between the target and projectile nuclei. This implies that the electron cloud instantly adjusts to the ground state of a molecule-like system consisting of the two approaching nuclei with respective charges of $Z_0$ and $Z_1$. The (negative) screening potential, $U_{\mathrm{ad}}$, can then be approximated by the difference in electron binding energies,
\begin{equation}
\label{eq:ebind}
U_{\mathrm{ad}} \approx  B_e(Z_0) + B_e(Z_1) - B_e(Z_0 + Z_1)
\end{equation}
where $B_e(Z_0)$, $B_e(Z_1)$, and $B_e(Z_0 + Z_1)$ denote the (positive) total electron binding energies in the atoms with charges of $Z_0$, $Z_1$, and $Z_0 + Z_1$, respectively (see Eq.~(5) in Ref.~\cite{Assenbaum87}).

As already pointed out in Sec.~\ref{sec:intro}, the values of $|U_e|$ extracted from low-energy cross section data are, in most cases, significantly larger than those calculated using the adiabatic approximation, $|U_{\mathrm{ad}}|$, by about a factor of two. A tabulated comparison between values can be found, e.g., in Ref.~\cite{spitaleri}. 

\section{Electron screening for narrow resonances}\label{sec:narrow}
For an isolated narrow resonance, what is usually measured is not directly the cross section, but the integrated cross section over the energy region of the resonance. This quantity is referred to as the resonance strength and can be extracted in the laboratory from the measured thick-target resonance yield curve \cite{Iliadis_2015}. The resonance strength, $\omega\gamma$, is defined by
\begin{equation}
\omega \gamma \equiv \omega \frac{\Gamma_a \Gamma_b}{\Gamma}
\end{equation}
where $\Gamma_a$, $\Gamma_b$, and $\Gamma$ $=$ $\Gamma_a$ $+$ $\Gamma_b$ $+$ ... denote the energy-dependent partial widths of the incoming channel and the outgoing channel, respectively, and the total resonance width; $\omega$ $\equiv$ $(2J + 1)/[(2j_p+1)(2j_t+1)]$ is the statistical spin factor, with $J$, $j_p$, and $j_t$ representing the spins of the resonance, projectile, and target, respectively. The general form of the resonance electron screening correction factor can then be written as
\begin{equation}
\label{eq:frgen}
f_r \equiv \frac{\omega\gamma_{\mathrm{screen}}}{\omega\gamma_{\mathrm{unscreen}}} = \frac{\Gamma_a^\prime}{\Gamma_a} \frac{\Gamma_b^\prime}{\Gamma_b} \frac{\Gamma}{\Gamma^\prime}
\end{equation}
where the primed and unprimed quantities refer to the screened and unscreened widths, respectively. The meaning of a ``narrow resonance'' in the present context will be defined at the end of this section.

In resonant charged-particle reactions at sufficiently low bombarding energies, which are of main interest in nuclear astrophysics measurements, the entrance channel width is much smaller than the exit channel width, i.e., $\Gamma_a$ $\ll$ $\Gamma_b$. In this case, Eq.~(\ref{eq:frgen}) reduces to
\begin{equation}
\label{eq:fr}
f_r = \frac{\Gamma_a^\prime}{\Gamma_a} = \frac{P^\prime}{P} \approx \frac{T^\prime}{T}
\end{equation}
Here, it is assumed that the main energy dependence of the particle partial width, $\Gamma_a$, arises from the penetration factor, $P_\ell$ (see, e.g., Ref.~\cite{Iliadis_2015}), and the latter quantity is approximated by the barrier transmission coefficient, $T$.\footnote{The definition of the transmission coefficient usually contains the ratio of wave numbers to the left and right of the barrier, whereas the penetration factor does not \cite{Blatt,Evans}. However, the wave numbers are implicitly included in the WKB wave function normalizations \cite{Merzbacher}. Therefore, the energy dependencies of the transmission coefficient and the penetration factor for the same value of the orbital angular momentum should be nearly equal.} 

In the opposite case, $\Gamma_a$ $\gg$ $\Gamma_b$, the resonance electron screening correction factor reduces to $f_r$ $\approx$ $\Gamma_b^\prime/\Gamma_b$. If such a resonance decays by emission of a $\gamma$ ray or neutron, electron screening will only impact the value of $f_r$ through the weak energy dependence of $\Gamma_b$, with the result that $f_r$ $\approx$ $1$.  If the emitted particle is charged (e.g., a proton or $\alpha$ particle), its transmission through the screened barrier must be considered in addition (see Eq.~(\ref{eq:frgen})).

\begin{figure}[ht]
\includegraphics[width=1.0\linewidth]{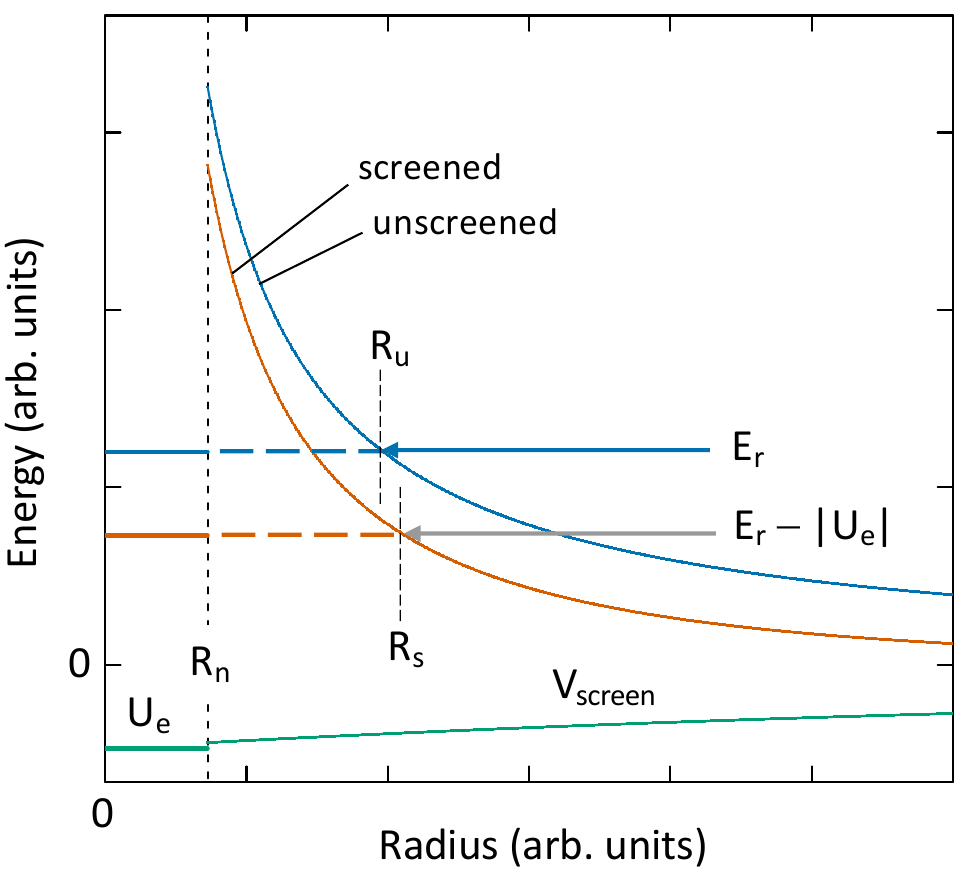}
\caption{\label{fig:res} 
Schematic representation (not to scale) of electron screening for a resonance in the laboratory, showing the unscreened Coulomb potential (blue curve), negative screening potential, $V_{\mathrm{screen}}$ (green), screened Coulomb potential (red curve), resonance energy, $E_r$ (blue arrow), and shifted energy, $E_r^\prime$ $=$ $E_r$ $-$ $|U_e|$ (gray arrow); $R_n$ denotes the nuclear radius, $R_u$ is the classical turning point at energy $E_r$ for the unscreened barrier, and $R_s$ is the turning point  at energy $E_r$ $-$ $|U_e|$ for the screened barrier. The actual reaction in the laboratory is represented by the gray arrow and the red curve. Notice that the tunneling distance, $R_s$ $-$ $R_n$, through the screened barrier at energy $E_r$ $-$ $|U_e|$ is larger than the distance $R_u$ $-$ $R_n$ through the unscreened barrier at $E_r$. If the screening potential, $V_{\mathrm{screen}}(r)$, would be constant, the tunneling distances would be the same and no change in either the transmission coefficient or resonance strength would be expected.
}
\end{figure}

Figure~\ref{fig:res} presents the situation for a  resonance with $\Gamma_a$ $\ll$ $\Gamma_b$, which is of primary interest in the present work. The unscreened Coulomb barrier is shown as the blue curve. The outer turning point for a particle approaching this barrier at the resonance energy, $E_r$, corresponding to a resonance level (blue horizontal line) inside the nucleus at the same energy, is denoted by $R_u$. The energy $E_r$ is a property of the compound nucleus only. Whereas outside the nuclear radius a constant screening potential was assumed for the discussion in Sec.~\ref{sec:nonres} and Fig.~\ref{fig:nonr}, this restriction will now be relaxed by adopting a negative screening potential, $V_{\mathrm{screen}}(r)$, that varies with distance (depicted in green in Fig.~\ref{fig:res}). At large radial distances, $r \rightarrow \infty$, the screening potential will approach zero, $V_{\mathrm{screen}}(r) \rightarrow 0$ (see also Sec.~\ref{sec:screen2}). Furthermore, inside the nucleus, the screening potential, $U_e$, is assumed to be constant (green horizontal line).\footnote{If we simplify the problem and assume that the K-shell electrons (see Sec.~\ref{sec:results}) form a uniformly charged sphere surrounding the target nucleus, then the screening potential will be nearly constant over the much smaller nuclear region. A constant screening potential inside the nucleus was also assumed, e.g., in Refs.~\cite{Zinner2007,cussons}.} 

A laboratory measurement of an isolated narrow resonance is impacted by electron screening in two ways: (i) outside the nucleus, the sum of the unscreened Coulomb potential (blue line) and screening potential (green line) gives rise to the screened Coulomb potential, shown in red; (ii) the attractive screening potential performs work on the projectile approaching the target atom, and, therefore, the energy at which the narrow resonance will be excited in the laboratory becomes $E_r^\prime$ $=$ $E_r$ $-$ $|U_e|$, where $E_r^\prime$ $<$ $E_r$ (see the gray arrow in Fig.~\ref{fig:res}). Or, expressed differently, the virtual level inside the compound nucleus (red horizontal line) is lowered by an amount of $|U_e|$. 

The transmission coefficent for the unscreened barrier is given by Eq.~(\ref{eq:nrscreen3}), where the center-of-mass resonance energy, $E_r$, replaces the energy, $E$. But, unlike the nonresonant case in Sec.~\ref{sec:nonres}, the transmission coefficient in the presence of electrons is given by
\begin{widetext}
\begin{equation}
\label{eq:tscr}
T^\prime \approx \exp \left( - \frac{ \sqrt{8 \mu}}{\hbar} \int_{R_n}^{R_s} \sqrt{ V_C(r) + V_{\mathrm{screen}}(r) - (E_r + U_e)} \, dr \right)
\end{equation}
\end{widetext}
where the outer turning point for the screened case, $R_s$, is obtained from $V_C(R_s)$ $+$ $V_{\mathrm{screen}}(R_s)$ $=$ $E_r$ $+$ $U_e$. It can be seen that, for the special case of a constant screening potential over the region of the outer turning point, i.e., $V_{\mathrm{screen}}(r)$ $=$ $U_e$ $=$~const, the two effects discussed above, (i) and (ii), cancel each other exactly. Consequently, the two turning points for the screened and unscreened case, $R_s$ and $R_u$, would coincide and Eq.~(\ref{eq:tscr}) reduces to Eq.~(\ref{eq:nrscreen3}). In other words, the electron screening correction factor, $f_r$, would become unity. This also means, contrary to the claim in Ref.~\cite{Assenbaum87}, that it is incorrect to apply the screening factor for nonresonant reactions, $f_{nr}$ in Eq.~(\ref{eq:nrscreen2}), to the measured strength of an isolated narrow resonance, because this procedure disregards the shift down in resonance energy from $E_r$ to $E_r$ $-$ $| U_e| $ in the calculation of the transmission coefficient. The possibility of measuring this resonance energy shift will be addressed in Sec.~\ref{sec:shift}.

When $V_{\mathrm{screen}}(r)$ is not constant, but declines outside the nuclear radius toward zero, the transmission coefficient for the screened Coulomb barrier is, in fact, {\it smaller} than the transmission through the unscreened barrier. This can be seen in Fig.~\ref{fig:res}, where the distance the particle needs to tunnel through the screened barrier, $R_s$ $-$ $R_n$, at $E_r$ $-$ $|U_e|$ is {\it larger} than the distance for tunneling through the unscreened barrier, $R_u$ $-$ $R_n$, at the energy $E_r$. Therefore, the unscreened resonance strength is generally larger than the screened value, which is the opposite of the assumption generally made in the literature for the laboratory screening correction for a narrow resonance (see Sec.~\ref{sec:intro}). In other words, unlike the correction factor for nonresonant cross sections, $f_{nr}$ $\ge$ $1$, the magnitude of the narrow-resonance correction factor is $f_r$ $\le$ $1$, as long as the screening potential, $V_{\mathrm{screen}}(r)$, is negative. It is assumed here that the screening potential, $U_e$, is constant inside the nucleus and can simply be subtracted from the unscreened resonance energy. It follows from the above discussion that the important quantity in this context is not only the magnitude of the electron screening potential, but also its rate of decline over the tunneling region.

Arguments similar to the above had been presented earlier in connection with electron screening in $\alpha$-particle radioactivity \cite{Zinner2007,Karpeshin} and screening effects for narrow resonances in astrophysical plasmas \cite{Mitler,cussons}. The shift in the energy of the virtual resonance level, caused by electron screening, is frequently disregarded in the literature (see, e.g., Refs.~\cite{erma,Liolios,Kettner_2006}), leading to incorrect predictions.

In the present context, a ``narrow resonance'' is defined by $\Gamma$ $\ll$ $|U_e|$, i.e., its total width must be small compared to the shift in the resonance energy, $U_e$ $=$ $E_r^\prime$ $-$ $E_r$, caused by electron screening. As discussed above, for this condition the reaction occurs at the screened energy, $E_r^\prime$, instead of the unscreened one, $E_r$. For a broad resonance, i.e., $\Gamma$ $\gtrsim$ $|U_e|$, the reaction can proceed over an extended range of incident energies, including the unscreened resonance energy, and the electron screening correction factor must be computed numerically from an expression more complicated than Eq.~(\ref{eq:fr}).  

\section{Screening lengths and screening potentials}\label{sec:screen2}
A simple model is used in this work to estimate numerical values for the screening effects on the measured strength of a narrow resonance. The resonance screening factor, $f_r$, is found by numerically integrating Eqs.~(\ref{eq:nrscreen3}) and (\ref{eq:tscr}). A Yukawa-type expression is adopted for the screened Coulomb potential outside the nuclear radius, 
\begin{equation}
\label{eq:yukawa}
V_C(r) + V_{\mathrm{screen}}(r) = \frac{ e^2}{4 \pi \epsilon_0} \frac{Z_0 Z_1}{ r}~e^{-r/ L} \,\, , r \ge R_n 
\end{equation}
where $L$ represents the electron screening length scale. The exponential factor damps the overall potential to nearly zero after a few screening lengths. For $r$ $\ll$ $L$, and keeping only the linear term in the expansion of the exponential factor, Eq.~(\ref{eq:yukawa}) reduces to
\begin{equation}
\label{eq:yukawa2}
V_C(r) + V_{\mathrm{screen}}(r) \approx \frac{ e^2}{4 \pi \epsilon_0} \frac{Z_0 Z_1}{ r} - \frac{ e^2}{4 \pi \epsilon_0} \frac{Z_0 Z_1}{ L} \,\, , r \ll L
\end{equation}
Therefore, and following Refs.~\cite{Zinner2007,cussons}, the constant screening potential inside the nucleus, $U_e$, can be approximated by 
\begin{equation}
\label{eq:du}
U_e =  - \frac{ e^2}{4 \pi \epsilon_0} \frac{Z_0 Z_1}{ L} 
\end{equation}
For the nuclear radius, a value of 
\begin{equation}
\label{eq:radius}
R_n = 1.2 (A_0^{1/3} + A_1^{1/3})~\mathrm{fm}
\end{equation}
will be assumed, where $A_0$ and $A_1$ are the integer mass numbers of the projectile and target, respectively. 

The last task before the electron screening factor for narrow resonances, $f_r$, can be computed numerically is to specify the electron screening length, $L$. The smaller the screening length scale, the larger the screening energy inside the nucleus, $U_e$, and its rate of decline outside the nuclear radius, and the larger the screening correction factor, $f_r$, will become. The screening length will depend on the atoms under consideration and the environment in which the nuclear reaction takes place.

A dominant contribution to the electron screening is provided by the (inner) core electrons, especially the K electrons. Their contribution will be estimated by approximating their screening length with the radius of the K shell,
\begin{equation}
\label{eq:core}
L_{KS} = r_{K} 
\end{equation}
The latter values were calculated by Ref.~\cite{Kohout} using the electron localization function (ELF), together with Hartree-Fock wave functions of the neutral atoms. Typical values of $r_{K}$ range from $0.58 a_0$ for carbon to $0.094 a_0$ for iron, where $a_0$ $=$ $5.29 \times 10^4$~fm denotes the Bohr radius.  

When the target atoms either form a metal lattice or are embedded in a metal backing, the screening effect  of the (free) conduction-band electrons must be considered in addition. An approximation of their screening length can be obtained from the Thomas-Fermi model of a metal \cite{Ashcroft}, which predicts\footnote{The numerical value of $3.7 \times 10^{-10}$ provided in Eq.~(3) of Ref.~\cite{Zinner2007} is incorrect and should be replaced by $6.1 \times 10^{-9}$.} 
\begin{equation}
\label{eq:TF}
L_{TF} = \sqrt{ \frac{2 \epsilon_0 E_F}{3 \rho e^2}} = 6.1 \times 10^{4} \sqrt{ \frac{E_F~[\mathrm{eV}]}{\rho~ [10^{22}~\mathrm{cm^{-3}} ] }  } ~ \mathrm{fm}
\end{equation}
where $E_F$ denotes the Fermi energy and $\rho$ is the electron density. Typical values for metals are $E_F$ $\approx$ $10$~eV and $\rho$ $\approx$ $10 \times 10^{22}$~cm$^{-3}$ \cite{Ashcroft}, giving a shielding length of $L_{TF}$ $\approx$ $6.10 \times 10^4$~fm. 

A number of authors (see, e.g., Refs.~\cite{bonomo,Raiola,Kettner_2006}) have computed screening lengths using the Debye-H\"uckel model, which yields\footnote{The numerical value of $2.18 \times 10^{-8}$ provided in Eq.~(4) of Ref.~\cite{Zinner2007} is incorrect and should be replaced by $2.18 \times 10^{-11}$.}
\begin{equation}
\label{eq:DH}
L_{DH} = \sqrt{ \frac{\epsilon_0 k_B T}{\rho e^2}} = 6.9 \times 10^{2} \sqrt{ \frac{T~[\mathrm{K}]}{\rho~ [10^{22}~\mathrm{cm^{-3}} ] }  } ~ \mathrm{fm}
\end{equation}
where $k_B$ and $T$ denote the Boltzmann constant and temperature, respectively. This model gives much smaller screening lengths, resulting in a stronger electron screening effect. Equation~(\ref{eq:DH}) is useful for a plasma \cite{Salpeter1954}, but this formulation does not apply to metals at room temperature, as pointed out, e.g., by Refs.~\cite{Huke2008,Dzyublik}. For doped semiconductors or electrolytes, the Debye-H\"uckel model results in modified expressions \cite{Ashcroft} compared to Eq.~(\ref{eq:DH}).

Here, only the dominant contributions to the electron screening, according to Eqs.~(\ref{eq:core}) and (\ref{eq:TF}), are considered. For a metal target and low bombarding energies, the velocity of the incident projectile is much smaller than the Fermi velocity of the electrons, and, therefore, the electron screening effect is caused by the static polarization of both the surrounding bound and conduction electrons. When applicable, the effects of K-shell and conduction electrons will be combined by adopting a shielding length of $L^{-1}$ $=$ $r_{K}^{-1}$ $+$ $L_{TF}^{-1}$, which assumes that the total screening potential is given by the sum of the individual contributions. Numerical results will be presented in the next section.

\section{Results and discussion}\label{sec:results}
Table~\ref{tab:results} gives the main results, including a comparison with values from the literature. Six narrow resonances are listed in the reactions $^{17}$O(p,$\alpha$)$^{14}$N, $^{18}$O(p,$\gamma$)$^{19}$F, $^{22}$Ne(p,$\gamma$)$^{23}$Na, $^{25}$Mg(p,$\gamma$)$^{26}$Al, and $^{27}$Al(p,$\gamma$)$^{28}$Si. All of these fulfill the conditions $\Gamma_a$ $\ll$ $\Gamma_b$ and $\Gamma$ $\lesssim$ $100$~eV (see Sec.~\ref{sec:narrow}). The target compositions are given in column~4. They range from wide-gap semiconductor material (Ta$_2$O$_5$), gas ($^{22}$Ne), to metal (Mg, Al). The screening lengths of the K-shell electrons in the neutral target atoms, $r_{K}$, which are listed in column~5, were assumed to be approximately equal to the K-shell radii found in Tab.~1 of Ref.~\cite{Kohout}. For the two metals, the screening lengths, $L_{TF}$, calculated from the Thomas-Fermi model according to Eq.~(\ref{eq:TF}), are given in column~6. The outer turning point radii, $R_s$, of the screened Coulomb potential, calculated from Eq.~(\ref{eq:tscr}), are listed in column~7. A comparison of length scales indicates that the screening lengths, $r_{K}$ and $L_{TF}$, are much larger than the outer turning-point radii, $R_s$. Consequently, any screening correction factors are expected to be small. Column~8 provides values for the constant screening potential, $U_e$ (see Eq.~(\ref{eq:du})), inside the compound nucleus, which are approximately equal to the energy difference between the unscreened resonance energy, $E_r$, and the screened one. Values of $U_e$ range from $-0.5$ to $-2.0$~keV. They are similar to the adiabatic approximation estimates obtained from Eq.~(\ref{eq:ebind}), which are given in column~9. 

The present estimates of the screening correction factors for narrow resonances, $f_r$, calculated according to Eqs.~(\ref{eq:fr}) $-$ (\ref{eq:TF}), are listed in column~10. As can be seen, the values of $f_r$ are unity within $0.2$\%. Also, the results predict that the screened resonance strengths are slightly {\it smaller} than the unscreened ones, consistent with the discussion in Sec.~\ref{sec:narrow}.

In comparison, screening ``enhancement'' factors for narrow resonances from the literature, $f_{\mathrm{Lit}}$, calculated from Eqs.~(\ref{eq:nrscreen5}) and (\ref{eq:ebind}), are given in column~11. These factors yield screened resonance strengths that {\it exceed} the unscreened values by 7\% to 25\%, depending on the reaction. Again, it must  be emphasized that it is not appropriate to calculate electron screening factors for narrow resonances using Eq.~(\ref{eq:nrscreen5}), which applies to nonresonant cross sections and disregards the shift in the resonance energy, as explained in Sec.~\ref{sec:narrow}. Notice, that the (incorrect) literature ``enhancement'' factors are significant, even when the measured resonance strength uncertainties are taken into account.

A number of tests were performed to investigate the sensitivity of the present results to parameter variations. Changing the nuclear radius parameter in Eq.~(\ref{eq:radius}) from $1.2$~fm to either $0.5$~fm or $2.0$~fm did not impact the numerical values of $f_r$ noticeably. The inclusion of a centrifugal term, $\hbar^2 \ell (\ell+1)/ (2 \mu r^2 )$, in Eqs.~(\ref{eq:nrscreen3}) and (\ref{eq:tscr}), and varying the orbital angular momentum, $\ell$, between $0$ and $3$, did not change any of the results either. Increasing the screening lengths adopted here (i.e., the values of $r_K$ and $L_{TF}$ listed in columns~5 and 6, respectively, of Tab.~\ref{tab:results}) will result in values of $f_r$ that are even closer to unity. When the screening lengths are reduced by a factor of two, the electron screening correction factors, $f_r$, are unity within 1\%. These changes are negligibly small, contrary to the correction factors reported in the literature for narrow resonances (column~11).

The simple procedure for calculating narrow-resonance screening factors presented here has a number of shortcomings. A static, time-independent model has been adopted, although a dynamical approach would be more appropriate. A constant screening potential is assumed inside the compound nucleus, see Eq.~(\ref{eq:du}), which oversimplifies the actual situation. Similar arguments apply to approximating the screened potential by the damped, Yukawa-type, expression of Eq.~(\ref{eq:yukawa}). The numerical results are impacted slightly by the adopted values of the screening lengths for the K-shell and conduction electrons, for which rough estimates have been employed here. It is worthwhile to address these issues in the future using more sophisticated models, e.g., similar to those developed for the related case of $\alpha$-particle radioactivity \cite{Patyk,Karpeshin,Dzyublik}. 


%
\begin{table*}[]
\begin{center}
\caption{Electron screening factors, $f_r$, and related quantities, for reported measured narrow resonances.}
\label{tab:results}
\begin{ruledtabular}
\begin{tabular}{lccccccccccc}
Reaction                            &  E$_r^{c.m.}$~(keV)   & $\Gamma$~(eV)              & Target                         & $r_{K}$~(fm)\footnotemark[1]   &   $L_{TF}$~(fm)\footnotemark[2]  &  $R_s$~(fm)\footnotemark[3]  &   $U_e$~(keV)\footnotemark[4]    & $U_{\mathrm{ad}}$~(keV)\footnotemark[5]   & $f_r$\footnotemark[6]       &  \multicolumn{2}{c}{Literature}   \\
\cmidrule(lr){11-12}
                                    &                      &             &                    &            &                         &                   &              &                       &              &  $f_{\mathrm{Lit}}$   &       Ref.   \\
\hline
$^{17}$O(p,$\alpha$)$^{14}$N        & 64.5                 & 130$\pm$5\footnotemark[8]  & Ta$_2$O$_5$\footnotemark[12]   &  21160                          &          &  178.6     &  $-$0.54   &   $-$0.68        &  0.9996      &  1.15           &                 \cite{PhysRevLett.117.142502}   \\
$^{18}$O(p,$\gamma$)$^{19}$F        & 90.0                 & 121$\pm$5\footnotemark[9]  & Ta$_2$O$_5$\footnotemark[12]   &  21160                          &          &  128.0     &  $-$0.54   &   $-$0.68        &  0.9998      &  1.10           &                \cite{Best2019}   \\
$^{18}$O(p,$\alpha$)$^{15}$N        & 90.0                 & 121$\pm$5\footnotemark[9]  & Ta$_2$O$_5$\footnotemark[12]   &  21160                          &          &  128.0     &  $-$0.54   &   $-$0.68        &  0.9998      &  1.09           &                 \cite{bruno,brunothesis}   \\
$^{22}$Ne(p,$\gamma$)$^{23}$Na      & 149.4                & $<60$\footnotemark[10]      & Ne gas                         &  15870                          &          &  96.4      &  $-$0.91   &  $-$0.91        &  0.9998      &  1.07           &                 \cite{PhysRevC.94.055804}    \\
$^{25}$Mg(p,$\gamma$)$^{26}$Al      & 92.2                 & $<30$\footnotemark[10]      & Mg metal                       &  12484                          &  55315   &  187.4     &  $-$1.7   &  $-$1.2        &  0.9976      &  1.25           &                 \cite{2012PhLB..707...60S}     \\
$^{27}$Al(p,$\gamma$)$^{28}$Si      & 956                  & 70$\pm$14\footnotemark[11] & Al metal                       &  11310                          &  49044   &  19.6      &  $-$2.0   &   $-$1.3       &  0.9999      &                 &                      \\
$^{176}$Lu(p,n)$^{176}$Hf           & 805\footnotemark[7]  &                            & Lu metal                       &                                 &          &            &            &           &              &                 &     \cite{Kettner_2006}     \\
\end{tabular}
\end{ruledtabular}
\footnotetext[1] {Electron screening length of K-shell electrons in the neutral target atom; see Tab.~1 of Ref.~\cite{Kohout}.}
\footnotetext[2] {Electron screening length of the Thomas-Fermi model for metals; see Eq.~(\ref{eq:TF}).}
\footnotetext[3] {Outer turning point of screened potential; see Eq.~(\ref{eq:tscr}).}
\footnotetext[4] {Constant screening potential inside the compound nucleus, which is approximately equal to the energy shift down from the unscreened resonance energy to the screened one; see Eq.~(\ref{eq:du}).}
\footnotetext[5] {Adiabatic approximation estimate of the screening potential; see Eq.~(\ref{eq:ebind}).}
\footnotetext[6] {Present estimate of the screening correction factor for narrow resonances; see Eq.~(\ref{eq:fr}).}
\footnotetext[7] {This resonance could not have been observed by Ref.~\cite{Kettner_2006} because its strength would be far below the present-day detection limit; see Sec.~\ref{sec:shift}.}
\footnotetext[8] {From Refs.~\cite{MAK198079,PhysRevC.71.055801}, using $\Gamma$ $\approx$ $\Gamma_{\alpha}$.}
\footnotetext[9] {From R matrix fit of Ref.~\cite{bruno} (see their Table 4).}
\footnotetext[10] {Upper limit using $\Gamma$ $\approx$ $\Gamma_{\gamma}$, with $\Gamma_{\gamma}$ estimated using the Recommended Upper Limits (RUL) \cite{ENDT1993171} for the primary transitions observed by Refs.~\cite{PhysRevC.94.055804} and \cite{PhysRevC.105.L042801} for $^{22}$Ne $+$ $p$ and $^{25}$Mg $+$ $p$, respectively.}
\footnotetext[11] {From Ref.~\cite{endt1990}.}
\footnotetext[12] {Wide-gap semiconductor.}
\end{center}
\end{table*}
%

\section{Resonance energy shifts caused by electron screening}\label{sec:shift}
Experimental studies of electron screening effects in resonant reactions face a number of obstacles. 

First, electron screening is expected to impact a resonance strength in a charged-particle induced reaction only when the entrance channel width is significantly smaller than the exit-channel one, $\Gamma_a$ $\ll$ $\Gamma_b$\footnote{For the condition $\Gamma_a$ $\gg$ $\Gamma_b$ (or $\omega\gamma$ $\approx$ $\omega \Gamma_b$), and assuming that the resonance decays by emission of a neutron or $\gamma$ ray, electron screening will impact the exit channel width, $\Gamma_b$, only through the small change in the decay energy (Sec.~\ref{sec:narrow}). In this case, the value of $f_r$ will be close to unity for an exothermic reaction.} (see Sec.~\ref{sec:narrow}). Second, even when the condition $\Gamma_a$ $\ll$ $\Gamma_b$ is fulfilled, the ratio of screened versus unscreened resonance strengths, $f_r$, will be close to unity (see Table~\ref{tab:results}) because the effects of the screened Coulomb potential and the shift in the resonance energy compensate each other largely (see Sec.~\ref{sec:screen2}). Consequently, electron screening will not significantly impact the values of measured resonance strengths, which are frequently extracted from the plateau height of thick-target yield curves \cite{Iliadis_2015}.

Because of these difficulties, it is worthwhile to consider, instead, to measure the shift in the resonance energy, $E_r^\prime$ $-$ $E_r$, caused by electron screening. Such a shift is expected to occur, in principle, in a charged-particle resonance reaction regardless of the relative magnitudes of the entrance and exit channel partial widths ($\Gamma_a$ and $\Gamma_b$).

A shift in resonance energy, presumably caused by electron screening, had been reported by Kettner et al.~\cite{Kettner_2006}. They measured the thick-target yield curve of a $^{176}$Lu(p,n)$^{176}$Hf resonance at 805~keV (center-of-mass energy), using three different target-backing combinations (Lu$_2$O$_3$ insulator, Lu metal, and PdLu alloy). No other information on this resonance is available in the literature. They observed an energy difference in the leading edge of the yield curves between the metal (and alloy) and the insulator target. By assuming that the insulator target exhibits insignificant screening, the observed energy shift down by $-32\pm2$~keV (Tab.~\ref{tab:results}) was interpreted as the electron screening potential for the metal (and alloy) target. Huke et al. \cite{Huke2008} discussed the energy shift reported by Ref.~\cite{Kettner_2006}, but attributed it instead to differences in target preparation and resulting stopping powers. The Wigner limit for an $s$-wave proton partial width in the $^{176}$Lu(p,n)$^{176}$Hf reaction at $805$~keV in the center of mass corresponds to a value of $\approx$ $10^{-16}$~eV, which is far below the present-day experimental detection limit. Therefore, the claim of Ref.~\cite{Kettner_2006} to have detected a resonance at $805$~keV in the $^{176}$Lu(p,n)$^{176}$Hf reaction, which is still being discussed in the recent literature \cite{2013EPJA...49...70G,2018EPJWC.16501035L,CVETINOVIC2023137684}, is incorrect.

No unambiguous evidence has so far been published demonstrating the existence of a shift in resonance energy caused by laboratory electron screening. Such an energy shift could be detected by comparing a resonance energy measured in the laboratory with the corresponding unscreened value. The latter corresponds to the resonance energy that would be obtained in the absence of all electrons surrounding the interacting nuclei. It can be determined from 
\begin{equation}
\label{eq:first}
E_r = E_x - Q_{\mathrm{nu}}
\end{equation}
where $E_r$ is the unscreened resonance energy in the center-of-mass system (same as Sec.~\ref{sec:narrow}), which is a property of the compound nucleus only; $E_x$ denotes the excitation energy of the compound nucleus, and $Q_{\mathrm{nu}}$ represents the $Q$ value computed from nuclear (as opposed to atomic) masses \cite{Iliadis:2019ch}. This value can be compared to the resonance energy that is obtained from a laboratory measurement, by using the relativistic expression 
\begin{equation}
\label{eq:second}
E_r^\prime = \sqrt{2 m_0c^2 E_r^{lab} + [(m_0 + m_1)c^2]^2} - (m_0 + m_1)c^2
\end{equation}
where $E_r^\prime$ and $E_r^{lab}$ denote the center-of-mass energy of the resonance in the presence of electrons (same as in Sec.~\ref{sec:narrow}), and the measured resonance energy in the laboratory reference frame, respectively; $m_0$ and $m_1$ represent the masses of the target and projectile. The energy shift caused by electron screening contributes to the measured difference, $E_r^\prime$ $-$ $E_r$.  

This procedure requires a careful assessment of all input quantities. The candidate resonance needs to be narrow (i.e., $\Gamma$ $\lesssim$ $100$~eV), and the target well characterized and free of surface contamination. The energy spread of the incident beam must be small (i.e., no more than a few hundreds of electron volts). The excitation energy, $E_x$, in Eq.~(\ref{eq:first}) needs to be precisely measured, preferably by $\gamma$-ray spectrometry. The laboratory resonance energy, $E_r^{lab}$, in Eq.~(\ref{eq:second}) must be measured precisely using methods that do not depend on the energies of other (calibration) resonances. Finally, additional effects caused by the presence of atomic electrons in the target need to be accounted for, e.g., the excitation and ionization of bound electrons in the atom in which the nuclear reaction is taking place \cite{Heinz_1987,PhysRevC.50.2466}, and the Lewis effect \cite{PhysRev.125.937}.

As an example, let us consider the resonance in the $^{27}$Al(p,$\gamma$)$^{28}$Si reaction near a center-of-mass energy of $956$-keV ($J^\pi$ $=$ $3^+$; $\Gamma$ $=$ $70\pm14$~eV \cite{endt1990}). The corresponding excitation energy, which was determined from the measured $\gamma$-ray energies of the primary decays, is reported as $E_x$ $=$ $12541.31\pm0.14$~keV \cite{ENDT1990209}. The nuclear $Q$ value amounts to $Q_{\mathrm{nu}}$ $=$ $11583.63\pm0.05$~keV \cite{wang2021}. Consequently, this yields an unscreened resonance energy of $E_r$ $=$ $957.68\pm0.15$~keV, according to Eq.~(\ref{eq:first}). The laboratory value of the resonance energy is reported as $E_r^{lab}$ $=$ $991.756\pm0.017$~keV \cite{BRINDHABAN1994436}. In that experiment, an aluminum metal target was used and the energy was determined relative to a Josephson-derived 1-V standard. Also, the reported value includes corrections caused by the ionization of atomic electrons (corresponding to an energy shift of $24\pm12$~eV). The above laboratory resonance energy results in a screened resonance energy in the center-of-mass system of $E_r^\prime$ $=$ $956.032\pm0.016$~keV, according to Eq.~(\ref{eq:second}). The energy difference, $E_r^\prime$ $-$ $E_r$, amounts to $-1.65\pm0.15$~keV. This result is near the screening energy of $U_e$ $=$ $-2.0$~keV (Table~\ref{tab:results}), which was estimated using the simple model of the present work, based on a Yukawa-type screening potential and screening lengths for electrons in the K shell and the conduction band (Sec.~\ref{sec:screen2}). It is also close to the value of $U_{\mathrm{ad}}$ $=$ $-1.3$~keV that is found from the adiabatic approximation (see Eq.~(\ref{eq:ebind})). Although these two estimates of the screening potential roughly agree with the energy difference, $E_r^\prime$ $-$ $E_r$, estimated above for the $E_r$ $=$ $956$-keV resonance in the $^{27}$Al(p,$\gamma$)$^{28}$Si reaction, further studies will be needed to confirm this claim. 

\section{Summary}\label{sec:summary}
The present work addressed the estimation of laboratory electron screening correction factors for isolated narrow resonances. Such corrections are frequently performed in the literature with the same procedure and expression used to correct laboratory nonresonant cross sections. It was pointed out that electron screening affects nonresonant cross sections and resonance strengths differently, and that it is not appropriate to correct measured resonance strengths using the same procedure and expression employed for the correction of measured nonresonant cross sections. The reported literature screening factors applied to narrow resonances result in unscreened resonance strengths that are {\it smaller}, by 7\% to 25\% depending on the reaction, than the measured (screened) ones. On the contrary, the present work demonstrated that unscreened resonance strengths are equal to the measured ones within $0.2$\%. This small correction is of no practical importance. Unless demonstrated otherwise, measured resonance strengths do not need to be corrected for laboratory electron screening effects. 

Since electron screening has a negligible impact on the strengths of narrow resonances, any attempts to study such effects by measuring the thick-target yield are futile. Instead, and regardless of the relative magnitudes of the entrance and exit channel partial widths, it may be more promising to detect the shift in the resonance energy down from the unscreened value (i.e., obtained in the absence of any electrons) to the screened one (i.e., measured in the laboratory). Although no unambiguous evidence has been published so far demonstrating such an energy shift, it is pointed out this effect is likely present in the data for the $E_r$ $=$ $956$-keV resonance in the $^{27}$Al(p,$\gamma$)$^{28}$Si reaction. It is also demonstrated that the claim of a previously reported detection \cite{Kettner_2006} of a resonance in the $^{176}$Lu(p,n)$^{176}$Hf reaction is incorrect.

%

\begin{acknowledgments}
The comments of Alain, Coc, Robert Janssens, Yosuke Kanai, Richard Longland, Caleb Marshall, and Thanassis Psaltis are highly appreciated. This work is supported by the DOE, Office of Science, Office of Nuclear Physics, under grants DE-FG02-97ER41041 (UNC) and DE-FG02-97ER41033 (TUNL). 
\end{acknowledgments}

\bibliography{paper}

\end{document}